\documentclass[aps,twocolumn,showpacs,preprintnumbers,pra]{revtex4-1}
\usepackage{graphicx,dcolumn,bm,amsmath,amssymb,hyperref,xcolor}
\usepackage[english]{babel}
\usepackage{siunitx}
\usepackage{bbm}
\usepackage{enumitem}  
\usepackage{float}
\newcommand{\n}[1]{{#1}}
\newcommand{\delete}[1]{{ }}

\newcommand{\pvec}[1]{\vec{#1}\mkern2mu\vphantom{#1}}

\renewcommand{\vec}[1]{{\mathbf{#1}}}

\begin{document}
\title{Skyrmion and Tetarton Lattices in Twisted Bilayer Graphene}
\author{\n{Thomas  B\"omerich} \href{https://orcid.org/0000-0001-9868-1497}{\raisebox{-0.1em}{\includegraphics[height=0.8em]{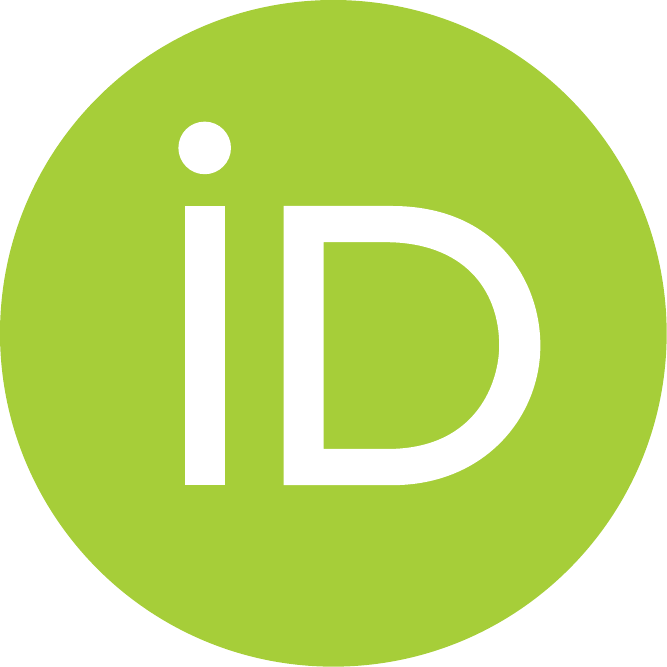}}}}
\author{Lukas Heinen \href{https://orcid.org/0000-0001-6638-7031}{\raisebox{-0.1em}{\includegraphics[height=0.8em]{orcid.pdf}}}}
\author{Achim Rosch \href{https://orcid.org/0000-0002-6586-5721}{\raisebox{-0.1em}{\includegraphics[height=0.8em]{orcid.pdf}}}}
\email{rosch@thp.uni-koeln.de}
\affiliation{Institute for Theoretical Physics, University of Cologne, D-50937 Cologne, Germany}
\date{\today}

\begin{abstract}
\noindent
Recent experiments on twisted bilayer graphene show an anomalous quantum
Hall (AQH) effect at filling of $3$ electrons per moir\`e unit cell. The
AQH effect arises in an insulating state with both valley- and
ferromagnetic order. We argue, that weak doping of such a system leads
to the formation of a novel topological spin texture, a `double-tetarton lattice'. The building block of this lattice, the `double-tetarton', is a spin configuration which covers $1/4$
of the unit-sphere twice. In contrast to skyrmion lattices, the net magnetization of this magnetic texture vanishes. Only at large magnetic fields more conventional skyrmion
lattices are recovered. But even for large fields the addition of a single charge to the
ferromagnetic AQH state flips hundreds of spins. Our analysis is based
on the investigation of an effective non-linear sigma model which includes the effects of
long-ranged Coulomb interactions.
\end{abstract}

\maketitle
\section{Introduction}
Twisted bilayer graphene (TBG) has emerged as a highly-tunable platform to observe correlated electron behaviour, such as insulating phases or unconventional superconductivity \cite{Cao2018,Cao_Insulator,Yankowitz2019}. By twisting two sheets of graphene by an angle $\theta$, a moir\'{e} pattern emerges and gives rise to a larger superlattice unit cell. The corresponding Brillouin zone (BZ) is much smaller than the BZ of a single graphene sheet and thus is called mini-BZ. At a `magic' twist angle of $\theta \approx 1.1 ^{\circ}$ the bands near the Fermi energy become exceptionally flat
\cite{Barticevic2010,CastroNeto2007,CastroNeto2012,SpectrumTBG,MacDonald2011}.
As the kinetic energy in these bands is small, electron-electron interactions become increasingly important. Because of spin and valley degeneracies the bands in the mini-BZ are four-fold degenerate. Besides controlling the bandstructure via the twist angle, the charge carrier density in twisted bilayer graphene can also be controlled by external electrostatic gating.\\
In the beginning of 2019, Sharpe \textit{et al}.\ found experimental evidence for a ferromagnetic state at filling $\nu=3$ \cite{Sharpe2019}. They measured an anomalous Hall effect which shows a hysteresis in an external magnetic field. There are many publications suggesting that the interactions may lift spin- and valley-degeneracies which could lead to different kinds of magnetic order \cite{MacDonald2020,Dodaro2018,Scheuer2018,Ochi2018,bultinck2019ground,Zaletel_skyrmions,Honerkamp2019}. Later in 2019, a quantized anomalous Hall effect was measured by Serlin \textit{et al}.\ in TBG on a hexagonal boron nitride (h-BN) substrate  for filling $\nu=3$ \cite{Serlin2019}. Because of the substrate the two-fold rotation symmetry of the TBG is broken, which gaps out the Dirac cones and the electronic bands aquire a non-zero Chern number $C$ \cite{GapsHBN,Jung2015,Zaletel2019,Zhang2019,Zhang_AQHE}. As the resulting ground state is a fully spin- and valley-polarized Chern insulator at filling $\nu=3$ \cite{Zaletel_skyrmions,liu2019correlated,alavirad2019ferromagnetism,repellin2019ferromagnetism,WuCollective2020}, there can be other charged excitations besides simple particle-hole pairs, namely skyrmions. 

In the quantum Hall phase there is a sizable Mott gap for charge excitations. Experimentally an activation gap of $ \SI{30}{K}$ has been measured in transport \cite{Serlin2019,Young2020}. Similarly, the valley degree of freedom is also gapped as its continuous rotation triggers a sign change of $\sigma_{xy}$ and thus closes the charge gap. Therefore, we do not expect topological textures involving the valley degree of freedom \cite{skyrmionQHRosch} and focus our study on the only remaining low-energy degree of freedom, the magnetization.
The spin structure can be described by a continuous vector field describing the classical magnetization 
$\hat{m}(\vec{r},t)$. A skyrmion has a non-trivial topology characterized by its winding number $W$:
\begin{equation}
W = \dfrac{1}{4 \pi}  \int \limits_{\mathbb{R}^2} \mathrm{d}^2 r\ \hat{m}(\vec{r})\cdot\left(\dfrac{\partial \hat{m}}{\partial x} \times \dfrac{\partial \hat{m}}{\partial y}\right) \in \mathbb{Z}  \label{eq:WindingNumber}
\end{equation}
If the Chern number of the electronic bands is independent of the spin orientation (as in the case of TBG),
the skyrmion aquires a charge given by the product of Chern- and winding number \cite{phaseSpaceBerry}.
Skyrmions are fermionic (bosonic) for odd (even) products. Interestingly, it has been argued \cite{khalaf2020charged} that superconductivity can arise from the condensation of bosonic skyrmions for $C=2$.

Long ago, it has been  established both theoretically \cite{Sondhi1993,Fertig1994,Fertig1997} and experimentally \cite{quantumHallFlippedSpin}, that spin-polarized electrons in the Landau levels of quantum Hall systems can form skyrmions which carry electric charge. The same is true for flat bands with a finite Chern number. The electric charge density $\rho_{\mathrm{el}}$ in this case is proportional to the topological winding density $\rho$:
\begin{equation}
\rho_{\mathrm{el}} = C e \rho \quad \text{with} \quad \rho =   \dfrac{1}{4\pi} \ \hat{m} \cdot \left( \dfrac{\partial \hat{m}}{\partial x} \times \dfrac{\partial \hat{m}}{\partial y}\right)  \label{eq:ChargeDensity}
\end{equation}

In the following we will numerically investigate topological textures induced by gating. Besides the expected skyrmion lattices we also find novel textures, which we dub
double-tetarton lattices. We study the phase diagram in a magnetic field and argue that a rapid change of magnetization as function of doping is a smoking guns signature of the double-tetarton phase.

\section{The model}

 The free energy of the magnetic sector can be described by a non-linear sigma model~\cite{Sondhi1993}:
\begin{align}
F[\hat{m}] =& \dfrac{J}{2}\int \limits_{\mathbb{R}^2} (\nabla \hat{m})^2 \ \mathrm{d}^2 \vec r -\int \limits_{\mathbb{R}^2} \vec{B} \cdot \hat{m}\ \mathrm{d}^2 \vec r  \label{eq:EnergyFunctional}  \\
& + \dfrac{U_c}{2} \int \limits_{\mathbb{R}^2} \int \limits_{\mathbb{R}^2}\dfrac{(\rho_{\mathrm{el}}(\vec{r})-\Delta \nu)(\rho_{\mathrm{el}}(\pvec{r}')-\Delta \nu)}{|\vec{r}-\pvec{r}'|}\ \mathrm{d}^2 \vec r \ \mathrm{d}^2 \vec r' \nonumber   
\end{align}
Here all lengths are measured in units of $L_M=\sqrt{A_M} \approx \SI{12}{nm}$, where $A_M$ is the area of the 
moir\'e unit cell \cite{Cao2018}. The first two terms describe the spin-stiffness $J$ of the ferromagnetic state and a Zeeman couling to an external field $B$. The third term is the long-ranged Coulomb interaction, $U_c=\frac{1}{4 \pi \epsilon_0 \epsilon L_M}$, between (topological) charges, where $\rho$ is the topological charge density defined in Eq.~\ \eqref{eq:ChargeDensity}.  $J,B,U_c$ have units of energy. $\Delta \nu$ is a background charge measured from filling $\nu=3$, which can be controlled by an external gate.
We assume that the distance to the gate is much larger than the average distance of charges. In this limit, the average charge density is fixed by $\Delta \nu$:
\begin{equation}
	\int \limits_{\mathbb{R}^2} \left( \rho(r)-\Delta \nu \right) \mathrm{d}^2r=0
\end{equation}
A scaling analysis, where all lengths are rescaled by the factor $\lambda$, reveals that the Coulomb energy and Zeeman energy scale with $\lambda^{-1}$ and $\lambda^2$, respectively, while the exchange term remains invariant. Coulomb repulsion (Zeeman energy) favors large (small) skyrmions. By minimizing the energy with respect to $\lambda$, one obtains an estimate for the radius of a single skyrmion in a magnetic field
\begin{equation}
R \sim L_M \left( \dfrac{U_c}{B} \right)^{1/3}=R^* \label{eq:Radius}
\end{equation}

\section{Groundstate at $B=0$} \label{Groundstate for B=0}
\begin{figure}[t]
	\centering
	\includegraphics[width=1.0\linewidth]{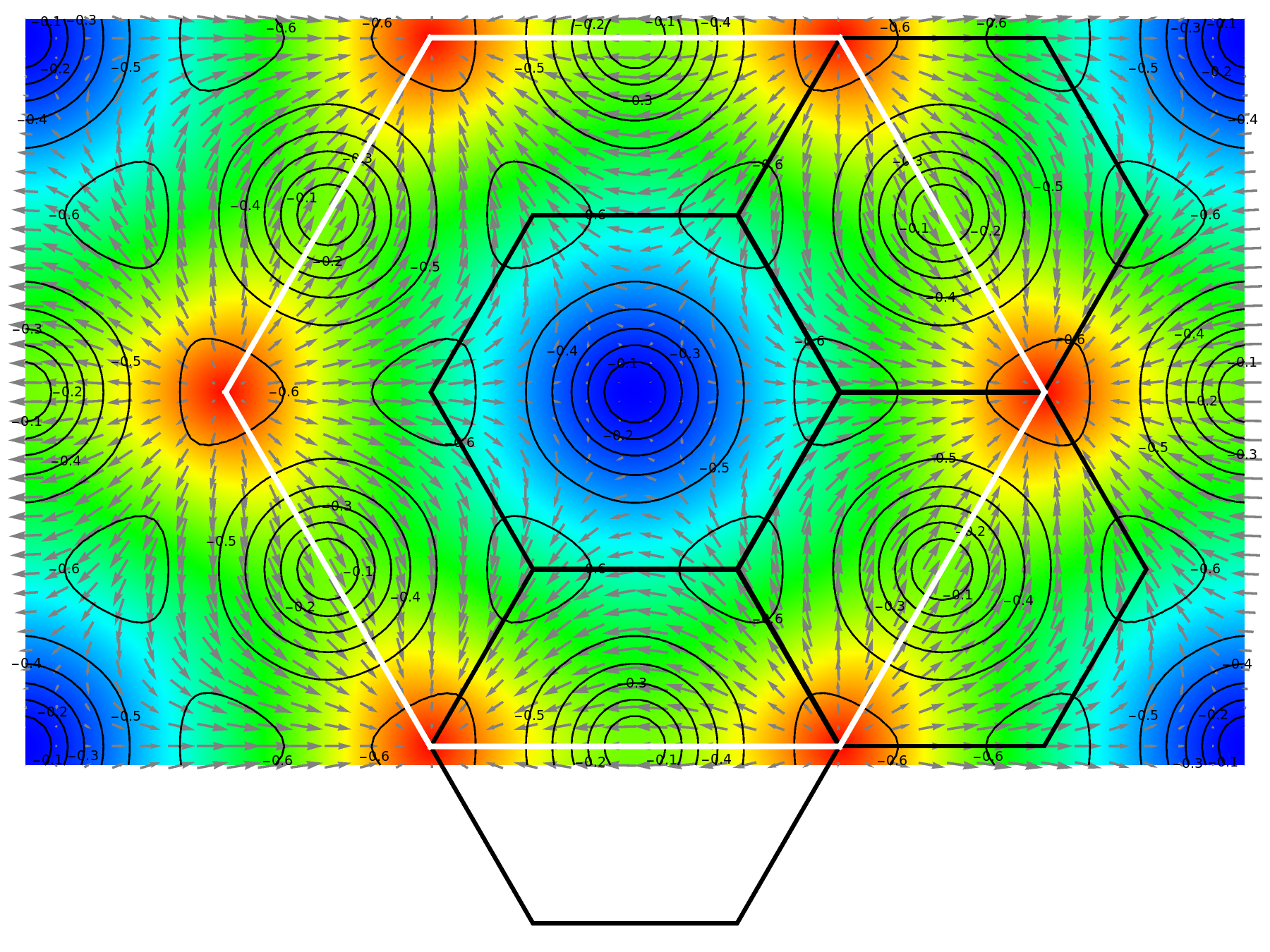}
	\caption{Groundstate spin configuration for $B=0$ and $\Delta \nu \left(\frac{U_c}{J}\right)^2
		=0.098$ (grey arrows: magnetization in the x-y-plane, colors: 
		z-component of the spin with blue for up and red for down spins). The total winding number within the 
		magnetic unit cell (white hexagon) is $W=-2$. The black hexagons depict the building blocks 
		of the magnetic structure, a `double-tetarton', see text and Fig.~\ref{fig:Tetron_cover}. The figure also shows contour lines
		of the topological charge density.}
	\label{fig:Groundstate_NoB}
\end{figure}

\begin{figure}[t]
	\centering
	\includegraphics[width=0.4 \linewidth]{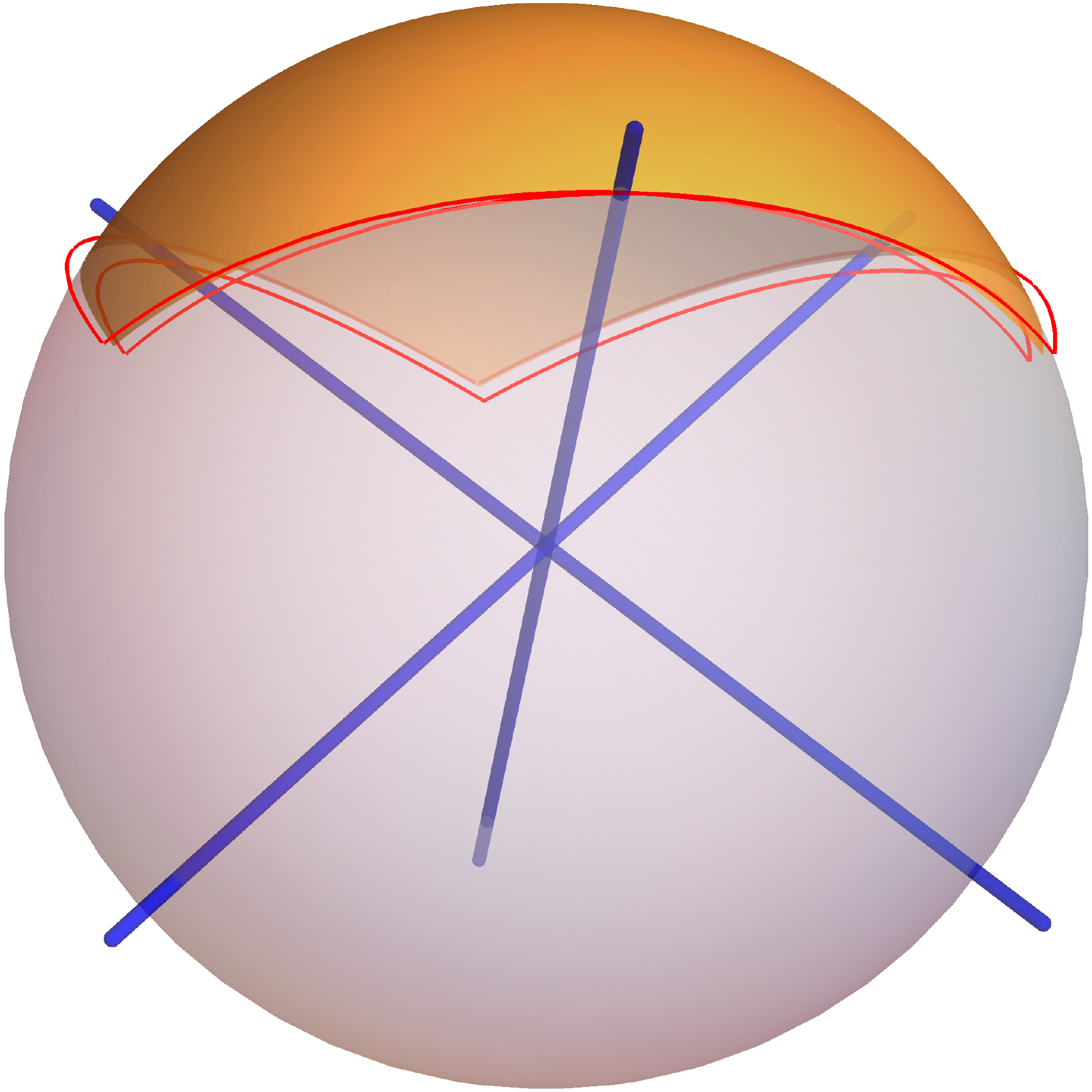}
	\caption{The basic building block of the texture shown in Fig.~\ref{fig:Groundstate_NoB} is a `double tetarton': 
		the spins cover exactly one quarter of the unit sphere twice. The red line shows a path along the edge of the central black hexagon which winds twice around the colored area. The magnetic texture of the other hexagons in Fig.~\ref{fig:Groundstate_NoB} can be obtained by rotating the spins by $180^\circ$ around one of the blue axes.}
	\label{fig:Tetron_cover}
\end{figure}

To determine the groundstate in the absence of a magnetic field at fixed winding number density, we performed numerical simulations for different unit cell geometries, see Appendix~\ref{simulations}. The lowest energy is found for 
a triangular lattice with a hexagonal unit cell shown in Fig.~\ref{fig:Groundstate_NoB}.
We first note that the total winding number within the magnetic unit cell, white hexagon in Fig.~\ref{fig:Groundstate_NoB}, is $-2$, but the resulting spin configuration is {\em not} 
a lattice of skyrmions.
The primary building block is instead the magnetic structure in the central black hexagon of Fig.~\ref{fig:Groundstate_NoB}. Here the spins cover exactly one quarter of the unit-sphere {\em twice} (a skyrmion covers  the full unit-sphere once). When one tracks the direction of spins moving along the edge of the central black hexagon,
one obtains a path shown in Fig.~\ref{fig:Tetron_cover} which winds twice around the northpole. In analogy to a `meron' (half of a skyrmion), we call this structure `double-tetarton'.

Moving from one black hexagon to the six next-nearest neighbours, the magnetic structure is rotated by 180$^\circ$
 around one of the three axes shown in Fig.~\ref{fig:Tetron_cover}. 
The group of magnetic symmetry transformations is -- up to the translations -- isomorphic to the octahedral group $O_h$, see Appendix~\ref{Symmetries}. Four double-tetartons thereby give the magnetic unit cell which therefore has 
winding number $W=4 \times 2 
\times \left(-\frac{1}{4}\right)=-2$.
By symmetry, the groundstate has no net magnetization 
$ \int  m_i(\vec{r}) \mathrm{d}^2r = 0$. This is an important observation which distinguishes our
double-tetarton lattice from skyrmion lattices. 
\begin{figure*}[t]
	\begin{center}
		\includegraphics[width=1.0\linewidth]{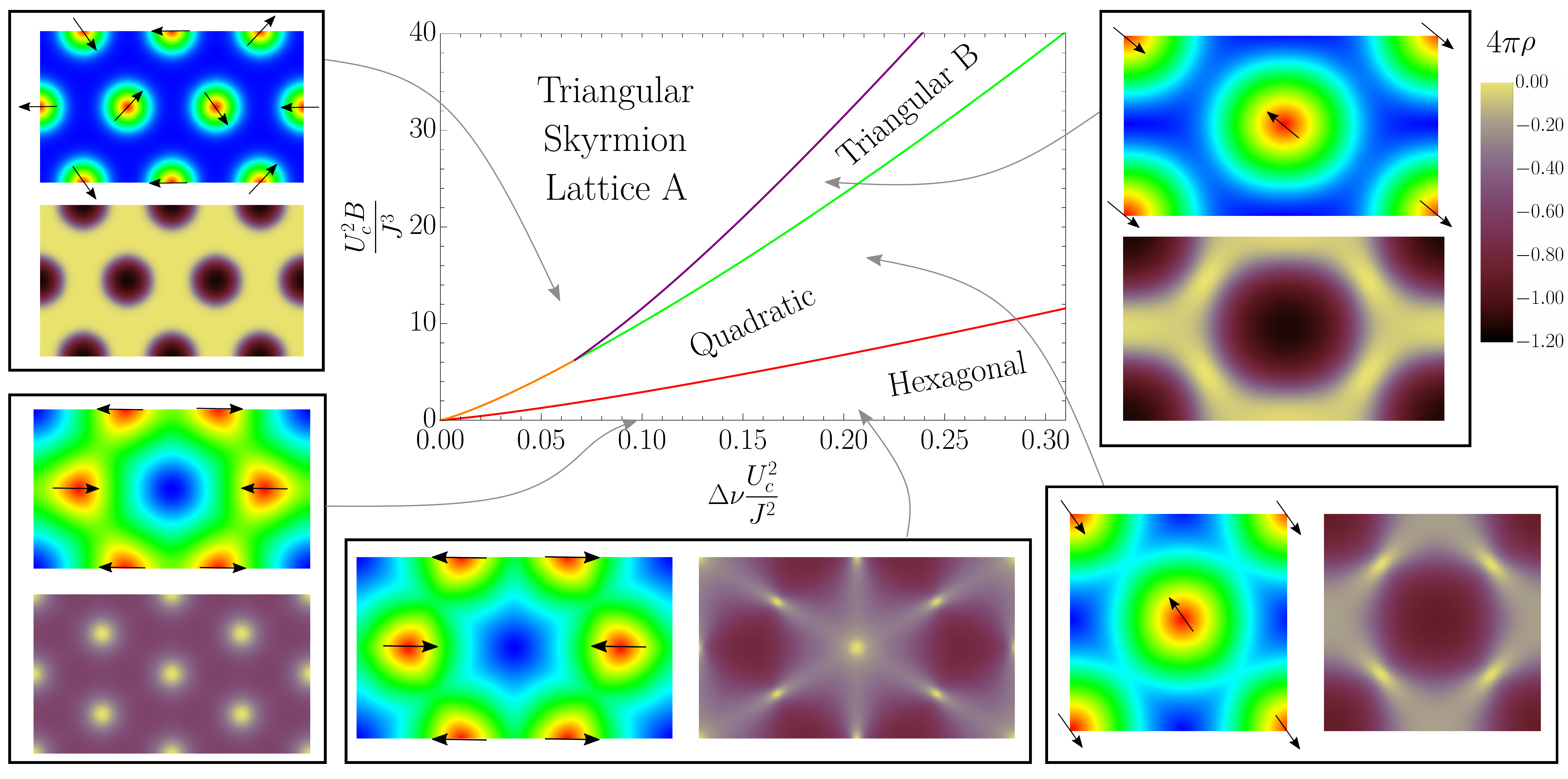}
	\end{center}
	\caption{Phase diagram for magnetic textures with Coulomb interactions. One representative spin configuration (color scale as in Fig.~\ref{fig:Groundstate_NoB}, arrows indicate the helicity) and the corresponding charge density for each phase is shown. In the case of $B=0$ the double-tetarton lattice (lower left corner) is the groundstate. For small magnetic fields a hexagonal lattice has the lowest energy (lower middle picture), while at low density and large magnetic field the groundstate is a triangular lattice of skyrmions with $120^\circ$ helicity order (upper left corner). At intermediate fields we obtain a triangular lattice with striped helicity order as well as a square lattice with an `antiferromagnetic' helicity order.}
	\label{fig:phaseDiagram}
\end{figure*}
Furthermore, we can look at the contours of the electric charge density depicted in Fig.~\ref{fig:Groundstate_NoB}.
The black hexagons define the unit cell of the charge density which has minima in their centers. 
The spin configuration spontaneously breaks global spin-rotation invariance and one can thus obtain other configurations just by rotating all spins, see Appendix~\ref{Symmetries}, 
the charge density remains invariant under such rotations.
In the limit $U_c \rightarrow 0$, when the energy is only determined by the exchange interaction, the energy $E_{UC}$ per magnetic unit-cell of the double-tetarton lattice is in the continuum limit exactly given by $E_{UC}=8 \pi J$,  twice the energy of the Polyakov skyrmion \cite{Polyakov}. 
This follows from the fact that the Polyakov skyrmion is a lower bound for the energy per winding number of topological textures in the presence of exchange interactions and that one can also construct an upper bound to the energy using lattices of Polyakov skyrmions. It is also consistent with our numerical results where we obtain for small $U_c$, $E_{UC} \approx 8 \pi J+ 0.04 \frac{U_c}{\sqrt{\Delta \nu}}$. If this energy is smaller than twice the Mott gap (the energy required to add two electrons into higher bands), then a topological magnetic texture will form whenever the system is doped slightly.

\section{Phase diagram}
In Fig.~\ref{fig:phaseDiagram} the phase diagram as a function of doping and 
magnetic field is shown.  A small magnetic field in z-direction breaks the $O(3)$ spin-rotation invariance. Numerically we find (see Appendix.~\ref{simulations}) that for a small magnetic field in the z-direction, the ground state smoothly evolves from the double-tetarton  configuration shown in Fig.~\ref{fig:Groundstate_NoB}. Due to the lowered symmetry, the double-tetarton lattice can now be smoothly deformed to a hexagonal lattice of skyrmions located at the six edges of the magnetic unit cell. Each skyrmion has an internal degree of freedom, called `helicity', which can be identified with the inplane-spin direction when moving from the skyrmion center in the $+\hat x$ direction. In the hexagonal small-field phase, the helicity (arrows in Fig.~\ref{fig:phaseDiagram}) shows an antiferromagnetic order.

In the opposite limit of large magnetic fields and small densities, the groundstate is given by magnetic skyrmions in a ferromagnetic background. These skyrmions are small and far apart from each other, so we can treat them as point-like particles which interact via Coulomb interactions. To minimize Coulomb repulsions,
they form a triangular lattice.  For large skyrmion distance, the helicity forms a $120^\circ$ order (triangular phase A), reminiscent of the magnetic order of antiferromagnetically coupled spins on triangular lattices. Indeed the helicities of neighbouring skyrmions are weakly (exponentially suppressed in the skyrmion distance) antiferromagnetically coupled via the ferromagnetic exchange interaction of spins. When the skyrmion radius $R \sim (U_c/B)^{1/3}$ becomes of the same order as the skyrmion distance $\sim 1/\sqrt{\Delta \nu}$, i.e., for $B\sim U_c (\Delta \nu)^{3/2}$, the skyrmions deform and helicity order changes to a striped state with opposite helicities (triangular phase B). Furthermore, we also obtain a centered square lattice between the hexagonal phase and the triangular skyrmion phases, see Fig.~\ref{fig:Groundstate_NoB}. In this phase the skyrmions show antiferromagnetic helicity order.

For an order-of-magnitude estimate of experimental parameters we assume $J \sim \SI{10}{meV}$ (of the same order of magnitude as the bandwidth) and $U_c=\frac{e^2}{4 \pi \epsilon_0 L_M}\sim \SI{100}{meV}$ (assuming $\epsilon \sim 1$). In our units  a magnetic field of one Tesla is equivalent to $B = \SI{0.06}{meV}$. The triple point in the phase diagram Fig.~\ref{fig:phaseDiagram}, where the triangular and the quadratic phases meet, is therefore predicted to occur at a doping of $\Delta \nu \approx 0.066 (J/U_c)^2 \sim 10^{-3}$ and a field of $B \approx 6.3 \frac{J^3}{\mu_B U_c^2} \sim \SI{10}{T}$. For a larger doping of a few percent, we expect that the system remains in the hexagonal phase for all experimentally accessible fields.

\section{Magnetization}
A central experimental signature \cite{quantumHallFlippedSpin} 
is the dependence of the magnetization per spin, 
$m_z$, on the charge or, equivalently, the skyrmion density. 
\begin{figure}[t]
	\centering
	\includegraphics[width=1.0\linewidth]{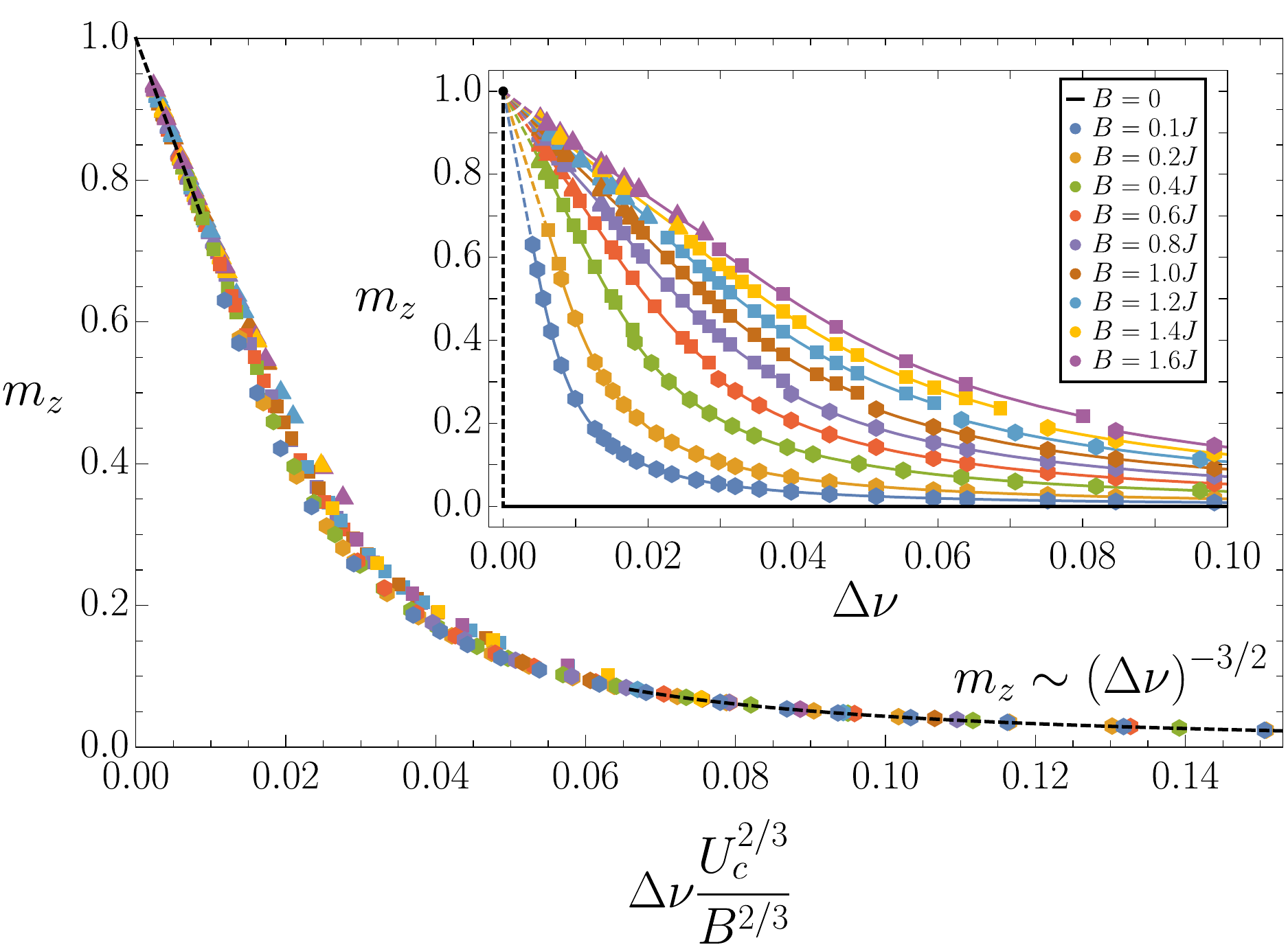}
	\caption{Magnetization (in units of $\mu_B$ per TBG moir\'e unit cell) as a function of the (rescaled) charge density. 
	Inset: For $B=0$ the magnetization jumps
	to zero for infinitesimal doping. The jump is broadened at finite $B$. 
	The shape of the markers (triangular, quadratic or hexagonal) indicate the 
	magnetic phases (triangular, quadratic or hexagonal). Note that there are tiny jumps
	at the first order transitions between two phases. Curves are taken for $U_c = 5J$.}
	\label{fig:magnetizationFitted}
\end{figure} 
For $B=0$ the ground state with finite winding number has  zero net magnetization as discussed above.
This implies that at $T=0$ and for $B \to 0$, the magnetization jumps from a fully polarized state, $m_z=1$,
to a state with zero magnetization for an arbitrarily small doping, $|\Delta \nu|>0$!
The inset of Fig.~\ref{fig:magnetizationFitted} shows that at finite $B$ field this jump is broadened 
to a crossover. For small $\Delta \nu$ the magnetization per skyrmion is given by
\begin{align}
M_{\rm sky} \approx 28 \left(\frac{U_c}{B}\right)^{2/3}
\sim {R^*}^2 \quad  \text{for } \Delta \nu\to 0, \label{eq:deltaM}
\end{align} 
which diverges for $B\to 0$, consistent with Eq.~\ref{eq:Radius}. 

This result suggests that the magnetization $m_z$  is a function of 
$\Delta \nu \left(\frac{U_c}{B}\right)^{2/3}$, 
\begin{equation}
m_z \approx f\left(\Delta \nu\left( \dfrac{U_c}{B}\right)^{2/3}\right)\label{scaling}
\end{equation}
 which is confirmed by the scaling plot of
 Fig.~\ref{fig:magnetizationFitted}. Note that we obtain only tiny jumps in the magnetization
when one crosses one of the first order transitions of Fig.~\ref{fig:phaseDiagram} and the magnetization of all
phases is  approximately described by the same scaling curve. 
For $B \to 0$, $m_z$ is linear in $B$ and therefore the scaling ansatz \eqref{scaling} predicts
 $f(x\to \infty) \sim x^{-3/2}$ or $m_z \sim (\Delta \nu)^{-3/2} B/U_c$ in this limit.

Our analysis has ignored the effects of dipolar interactions. Remarkably, simple power counting arguments show
that dipolar interactions should become important in the limit of infinitesimal doping  $\Delta \nu \to 0$. However,
an analysis of the relevant prefactors shows, that for realistic parameters the effects of dipolar interactions
are negligible, see Appendix~\ref{Dipolar}.

\section{discussion}
Twisted bilayer graphene provides a unique opportunity to discover new topological 
states of matter. Importantly, the anomalous quantum Hall effect in this system observed for $\nu=3$ is not induced
by spin-orbit coupling but arises from the ordering of the valley degree of freedom. Thus the spin degree of freedom
can rotate without closing the gap. We have argued that for small doping away from $\nu=3$, one therefore naturally realizes a topological magnetic texture with finite winding number and zero net magnetization best described as a lattice of `double tetartons', i.e., textures which cover $1/4$ of the unit sphere two times and which are connected to neighbouring tetartons by the three two-fold rotation axes of a tetrahedron. 

Experimentally, the most direct way to measure topological textures in twisted bilayer graphene is to use spin-polarized scanning tunneling microscopy. Also measurements of the magnetization as a function of the gate voltage can be used: whenever the number of flipped spins per added charge is large (according to Eq.~\ref{eq:deltaM} about
300 spins flip in a field of $\SI{10}{T}$), this clearly indicates the presence 
of skyrmionic excitations. As the double-tetarton lattice carries zero magnetization, we predict that this number diverges in the low-temperature, low-field limit.

An interesting  question is whether tetartons can exist as single particles. Here it is useful to consider the analogy with merons, half-skyrmions which cover $1/2$ of the unit sphere. They are realized in two-dimensional ferromagnets with an easy-plane anisotropy as vortex states \cite{MeronsAnisotropy,MeronsEasyPlane}. 
Similarly, we have checked that tetartons covering exactly $1/4$ of the unit sphere naturally arise
in two-dimensional ferromagnets with certain cubic anisotropies when, for example, 
three domains with orientation $(1,1,1)$, $(1,-1,-1)$ and $(-1,1,-1)$ meet. In  (anomalous) quantum Hall systems 
with Chern number $1$ such textures naturally carry the charge $1/4$.
For the future it will be interesting to investigate how such topological textures can be controlled by currents and fields and to explore possible classical and quantum liquids generated from such states.

\acknowledgements

The  numerical simulations have been performed with the open-source micromagnetic simulation program MuMax3 \cite{MuMax3,MinimizeMuMax} with custom additions, see Appendix \ref{simulations}, on the CHEOPS cluster at the RRZK Cologne. We thank S. Ilani, A. Vishvanath, and M. Zirnbauer for useful discussions and the DFG for financial support (CRC1238, project number 277146847, subproject C02). We thank M. Antonoyiannakis and A. Melikyan for suggesting the name tetarton.

\appendix

\section{Simulations} \label{simulations}
 Our simulation system is set up by choosing a rectangular unit cell and attaching copies of it in both space directions. This unit cell contains $N_x \cdot N_y$ lattice points with lattice spacing $a_x,a_y$. 
 
The Coulomb interaction in Eq.~\eqref{eq:EnergyFunctional} in the main text is a highly nonlocal 6-spin interaction containing two integrals over space. Therefore it has to be implemented in an efficient way. Assuming that the charge density is constant inside the discretization cell of each lattice point, the discretized version of the Coulomb interaction can be written as:
 \begin{equation}
 F_c [\hat{m}] = a_x^2\  a_y^2\ \dfrac{U_c}{2} \sum_{r,r'} \tilde{\rho}_r \tilde{\rho}_{r'} K_{r-r'} \quad \text{with} \quad \tilde{\rho}_r=\rho_r - \Delta \nu
 \end{equation}
where the sums run over all lattice points and the Coulomb kernel $K_{r-r'}$ given by
 \begin{equation}
 K_{r-r'} = \dfrac{1}{a_x^2 a_y^2}\int \limits_{V_r} \mathrm{d}^2 x \int \limits_{V_{r'}} \mathrm{d}^2 x' \dfrac{1}{|\pvec{x}-\pvec{x}'|}
 \end{equation}
 Here, the integrals are taken over the discretization cells $V_r$ and $V_{r'}$. Due to the singular nature of the Coulomb interaction, it is important to use these integrals instead of $1/|r_i - r_j|$.
 	We can write the free energy in terms of a potential $\phi_r$ acting on the charge $\tilde{\rho}_r$. The correction to the effective magnetic field acting on spin $i$ at site $\vec r_i$, which arises from the Coulomb interaction, is then computed from
 \begin{equation}
 \vec B^c_i = -\frac{\delta F_c [\hat{m}]}{\delta \hat m_i}= - U_c a_x a_y  \sum_{r} \frac{\delta \tilde{\rho}_r}{\delta \hat m_i} \phi_r 
 \label{Bc}
 \end{equation}
 
 with
 \begin{equation}
 \phi_r=a_x  a_y \sum_{r'} \tilde{\rho}_{r'} K_{r-r'}
 \end{equation}

 	To implement these equations in MuMax we proceed in the following way (roughly following the scheme how dipolar interactions are implemented in MuMax). First, upon start of the program, the matrix $K_{r-r'}$ is computed for the given discretization of the 2d plane using numerical integration 
 	(adaptive 2d-Gauss-Kronrod). We need the Fourier transform of this matrix, taking into account that we describe a periodic system. Thus the Fourier transform is defined as $K_k=\frac{a_x a_y}{N_x N_y} \sum \limits_{r,r',n_x,n_y} K_{r-r'-n_x L_x-n_y L_y} e^{-i \vec k (\vec r - \vec r')}$. This Fourier transformation is stored.
 
To compute the topological charge density $\tilde \rho$, we use the 2nd-order discretization of the topological charge, Eq.~\eqref{eq:ChargeDensity}, of the main text, as implemented in MuMax. We have used this formula to compute $\frac{\delta \tilde{\rho}_r}{\delta \hat m_i}$. To compute $\phi_r$ efficiently, we use fast Fourier transformations (FFT), recycling the code used by MuMax to compute dipolar interactions.
 	With these results, we can directly compute the effective magnetic field of Eq.~\eqref{Bc}, which enters directly the solution of the LLG equation within MuMax.

\begin{figure}[t]
	\centering
	\includegraphics[width=1.0\linewidth]{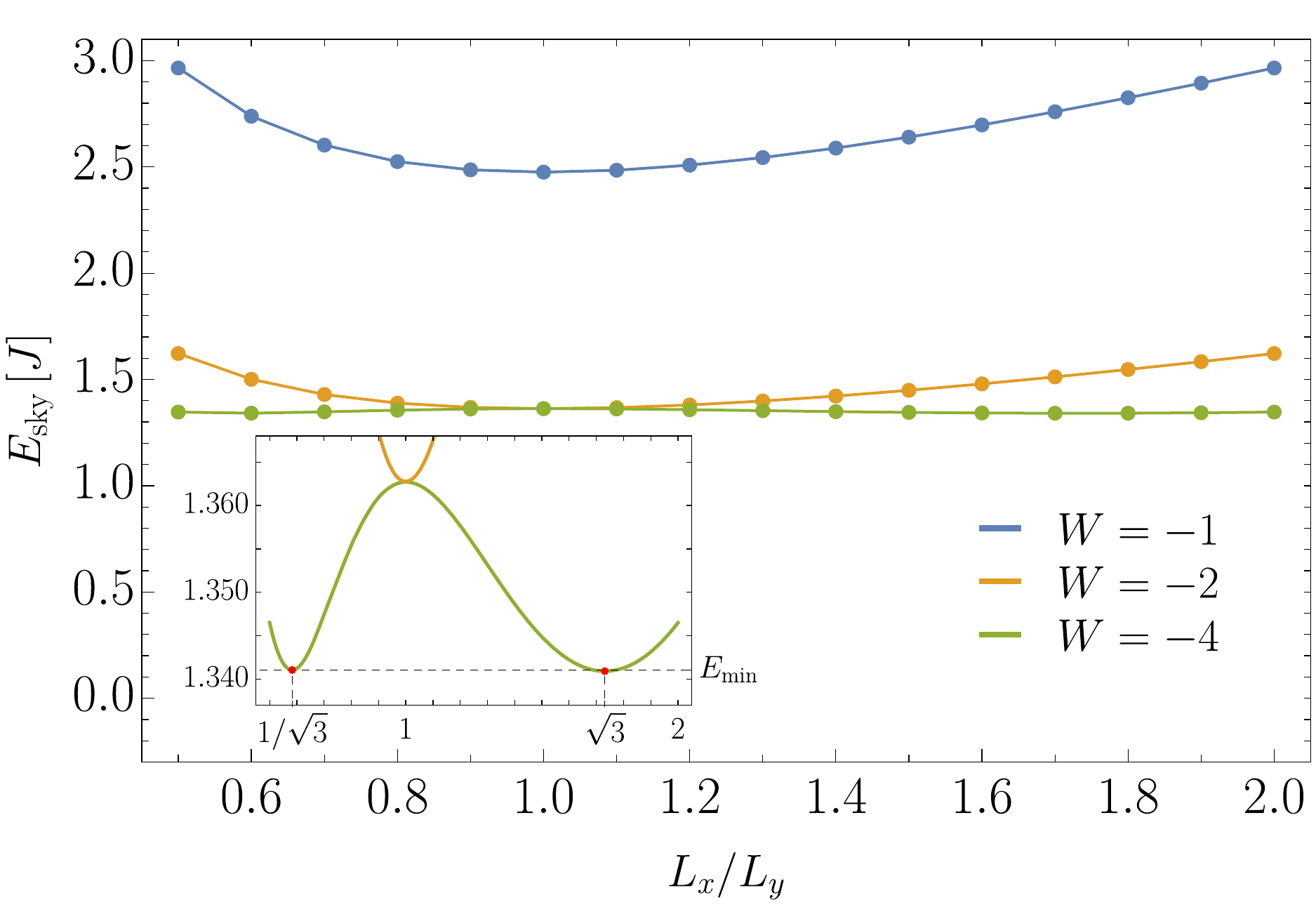}
	\caption{Energy density as a function of aspect ratio $L_x/L_y$ for different total winding numbers $W$ in the simulation cell ($U_c = 5J,\ B=0$). The inset shows the groundstate at $L_x/L_y$= $\sqrt{3}$ and $W=-4$ with spin configuration \ref{fig:Groundstate_NoB}.}
	\label{fig:GroundstateLengthsInset}
\end{figure}

 We can compare configurations with different winding numbers $W$ in the unit cell when we rescale the linear extents of our systems in such a way that the winding density is kept fixed. To identify the spin configuration with the lowest energy we start from various regularly arranged configurations of skyrmions and minimize the energy using a conjugate gradient method \cite{MuMax3,MinimizeMuMax}. Furthermore, we also looked for the energy minimum of randomly placed skyrmions in a large simulation cell. It turns out that for $B=0$ the lowest energy is obtained for $W=-4$ and a unit cell aspect ratio of $L_x/L_y=\sqrt{3}$, see Fig.~\ref{fig:GroundstateLengthsInset}. The second minimum arises from relabeling $L_x,L_y$.

Applying a magnetic field in z-direction lifts the degeneracy of our groundstate, as spin-rotation symmetry around x- and y-direction is lost. To find the right groundstate for a small magnetic field we take the spin configuration \ref{fig:Groundstate_NoB} and rotate all spins around the x-direction by an angle $\alpha_x$. Then we turn on a small magnetic field and minimize the energy of the rotated magnetization configuration and the unrotated one. The energy differences in this case are all very small but using this method we do not find any state with lower energy than the spin configuration arising from Fig.~\ref{fig:Groundstate_NoB}. The same can be done for a spin rotation around the y-direction and for different magnetic field strengths, which gives the same qualitative results.

\section{Symmetry group of the groundstate for $B=0$} \label{Symmetries}
The magnetic ground state is characterized by an unusual non-symmorphic symmetry group. While the charge density in Fig.~\ref{fig:Groundstate_NoB} has a $60^{\circ}$ rotation symmetry around the center of the unit cell, the corresponding spin configuration only has a $120^{\circ}$ symmetry.

As our energy functional \eqref{eq:EnergyFunctional} in the absence of a magnetic field is invariant under rotations of  spin, rotation of space, and translations, an element $g_i$ of the groundstate symmetry group can be written as 
$g_i=\left( M_i, D_i, \vec{t}_i\right)$, where $M_i$ is a $3\mathrm{x}3$ matrix describing rotations of the magnetization, $D_i$ is a $2\mathrm{x}2$ space rotation or reflection matrix and $\vec{t}_i$ is a translation vector.
\begin{align}
\hat m(\vec r) \stackrel{g_i}{\longrightarrow} \hat m'(\vec r)=M_i \hat m(D_i^{-1}\vec r-\vec{t}_i)
\end{align}
\begin{figure}[t]
	\centering
	\includegraphics[width=1.0\linewidth]{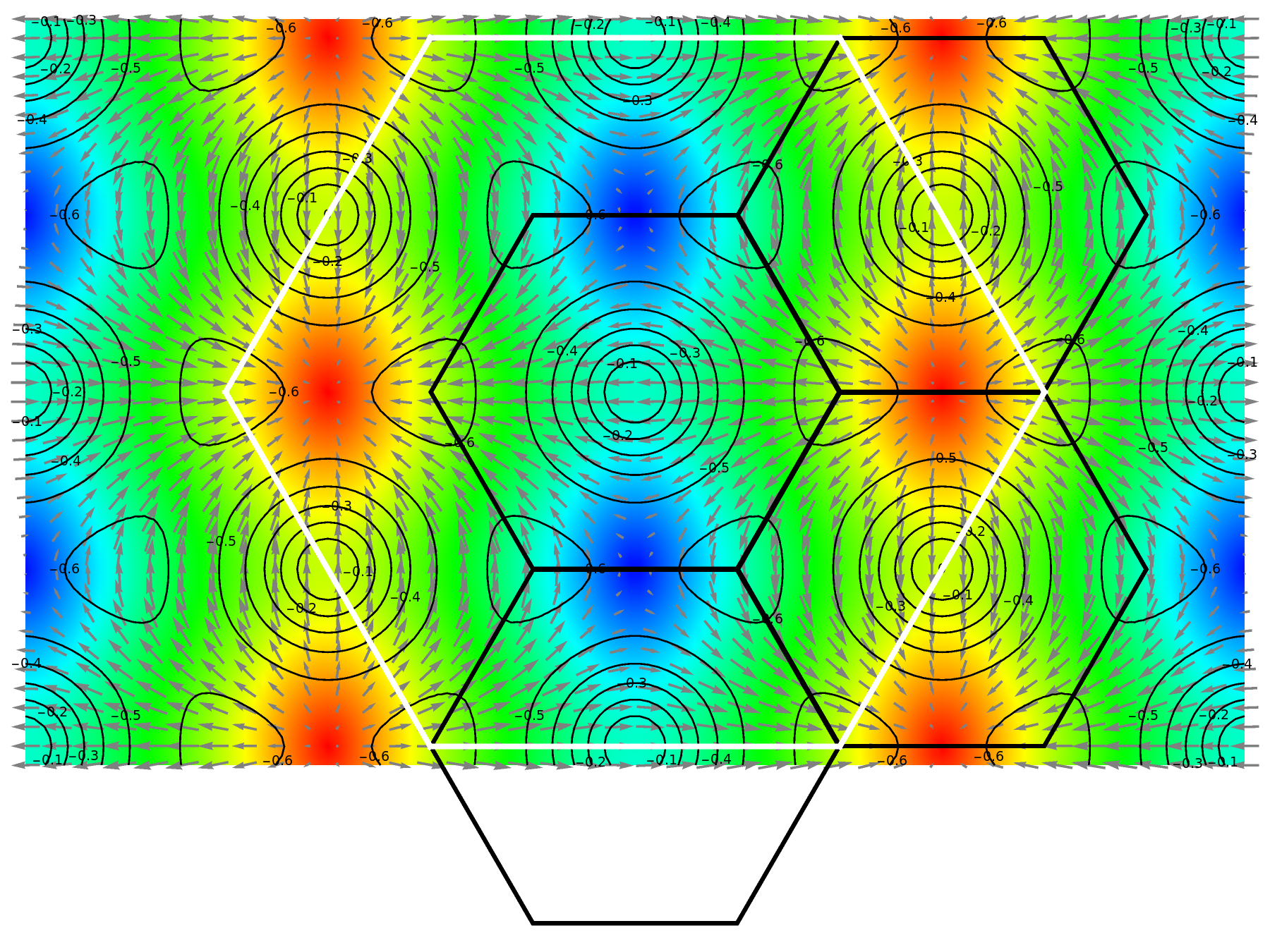}
	\caption{Spin configuration resulting from rotating all spins in Fig.~\ref{fig:Groundstate_NoB} by $\pi/3$ around the y-axis. The energy functional and the charge density are invariant under global spin-rotations.}
	\label{fig:AdditionalGroundstateHexagonal}
\end{figure}
The magnetic lattice is invariant under translations of a triangular lattice with translation vectors
$\vec T_1=(0,1)$ and $\vec T_2=(\sqrt{3},1)/2$. 
It is, however, also possible to combine a translation by $\vec T_1/2$ (by $\vec T_2/2$)
with a spin-rotation by $\pi$
around the axis $(-\sqrt{2},0,1)$ (around $\left( \frac{1}{\sqrt{2}},- \sqrt{\frac{3}{2}}, 1 \right)$) to a symmetry transformation. Denoting a rotation with the angle $\phi$ around the axis $\vec n$ by $M_\phi^{\vec n}$, we obtain the following group elements
\begin{enumerate}[label=(\roman*)]
	\item the identity: $g_0=\left(\mathbbm{1}_3,\mathbbm{1}_2,\vec{0} \right)$ 
	\item translation by half the lattice constant combined with spin rotations by $\pi$: \\[5pt]
	$g_1=\left(M_\pi^{(-\sqrt{2},0,1)},\mathbbm{1}_2,\vec T_1/2 \right)$ and \\[5pt] $g_2=\left(M_\pi^{\left( \frac{1}{\sqrt{2}}, -\sqrt{\frac{3}{2}}, 1 \right)},\mathbbm{1}_2,\vec T_2/2 \right)$ 
		\item rotation by $\frac{\pi}{3}$ around the z-axis combined with a spin-rotation by double the angle:\\[5pt]  
		$g_3=\left( M_{2 \pi/3}^{\hat z},D_{\pi/3}^{\hat z},\vec{0} \right) $
	\item space reflection, $x \to -x$, combined with a spin-rotation around the y-axis 
	by $\pi$ and time reversal:\\[5pt] $g_4=\left(  \left( \begin{array}{ccc} 
1 & 0 & 0 \\                                     
0 & -1 & 0 \\
0 & 0 & 1 \\                                              
\end{array}\right) , \left( \begin{array}{cc} 
	-1 & 0  \\                                     
	0 & 1  \\                                            
	\end{array}\right),\vec{0} \right) $
\end{enumerate}
In total the space group is spanned by translations by $\vec T_1$ and $\vec T_2$ combined with 
48 group operations (isomorphic to the octahedral group $O_h$) obtained from products of $g_0, g_1, \dots, g_4$.
For the discussion in the main text the symmetry operations, $g_1$, $g_2$ and $g_1 g_2$ (depicted by the 3 axes in Fig.~\ref{fig:Tetron_cover}) are most important, 
as they map the spin configuration of one black hexagon (one `double tetarton') of Fig.~\ref{fig:Groundstate_NoB} to the neighbouring hexagon. These symmetries also guarantee that the total magnetization of the texture vanishes.

As the topological charge density and the energy functional are spin-rotation invariant in the absence of a magnetic field, one can obtain different magnetic configurations with the same energy and charge density by spin rotations of the groundstate in Fig.~\ref{fig:Groundstate_NoB}. One example is shown in Fig.~\ref{fig:AdditionalGroundstateHexagonal}, where all spins are rotated by $\frac{\pi}{3}$ around the y-axis w.r.t.\ the spin configuration in Fig.~\ref{fig:Groundstate_NoB}.

\section{Dipolar interactions} \label{Dipolar}

Up to now we have neglected dipolar interactions between the spins. As these dipole-dipole interactions are also long-ranged, it is important to estimate the strength of this interaction compared to the Coulomb energy. The free energy for a discrete set of spins is given by:
\begin{equation}
F_{DD} = - \dfrac{\mu_B^2 \mu_0}{4 \pi} \sum \limits_{i<j} \dfrac{3 \left(\vec{\hat{r}}_{ij} \cdot \vec{s}_i \right)\left( \vec{\hat{r}}_{ij} \cdot \vec{s}_j \right) -\vec{s}_i \cdot \vec{s}_j}{|\vec{r}_{ij}|^3} \\
\end{equation}
where $\vec{r}_{ij}$ is the vector connecting the sites at $i$ and $j$ and $\mu_B$ is the magnetic moment of one spin. In the following, we will compare the short-ranged contribution $E_{\mathrm{SR}}$ of the dipolar energy coming from a region of size $L_R$ with the corresponding Coulomb energy $E_C$. In 2d systems, $E_{\mathrm{SR}}$ dominates the dipolar contribution and leads to an uniaxial anisotropy which wants to align the spins in-plane. An order of magnitude estimation of these two energies is given by:
\begin{equation}
 E_{\mathrm{SR}} \sim \dfrac{\mu_B^2 \mu_0}{4 \pi a_M^3} \left( \dfrac{L_R}{a_M}\right)^2 
\end{equation}
\begin{equation}
 E_C \sim \dfrac{e^2}{4 \pi \epsilon_0} \dfrac{1}{L_R}
\end{equation}
Here, $a_M$ is the discretization length of the spins which is given by the moir\'{e} distance as there is one electron per moir\'{e} unit cell. From these two expressions we can estimate the length scale $L_R$ for which dipolar interactions become more important than the Coulomb interaction:
\begin{equation}
	L_R \sim a_M \left( \dfrac{1}{\alpha} \dfrac{a_M}{a_B} \right)^{2/3} \sim 10^3 \ a_M
\end{equation}
where $\alpha \approx 1/137$ is the fine structure constant and $a_B \sim a_M/240$ is the Bohr radius. We can convert this length scale into a relative change of the electric charge density:
\begin{equation}
	\dfrac{\delta n}{n} \lesssim \left( \dfrac{a_M}{L_R} \right)^2 \sim 10^{-6}
\end{equation}
From this we can conclude that dipole-dipole interactions can be neglected except for really small doping. We have also investigated the long-ranged part of the dipolar interactions, which gives an even smaller contribution compared to the Coulomb interaction energy.

\section{Skyrmion radius} \label{skyrmionRadius}

\begin{figure}[b]
	\centering
	\includegraphics[width=0.9\linewidth]{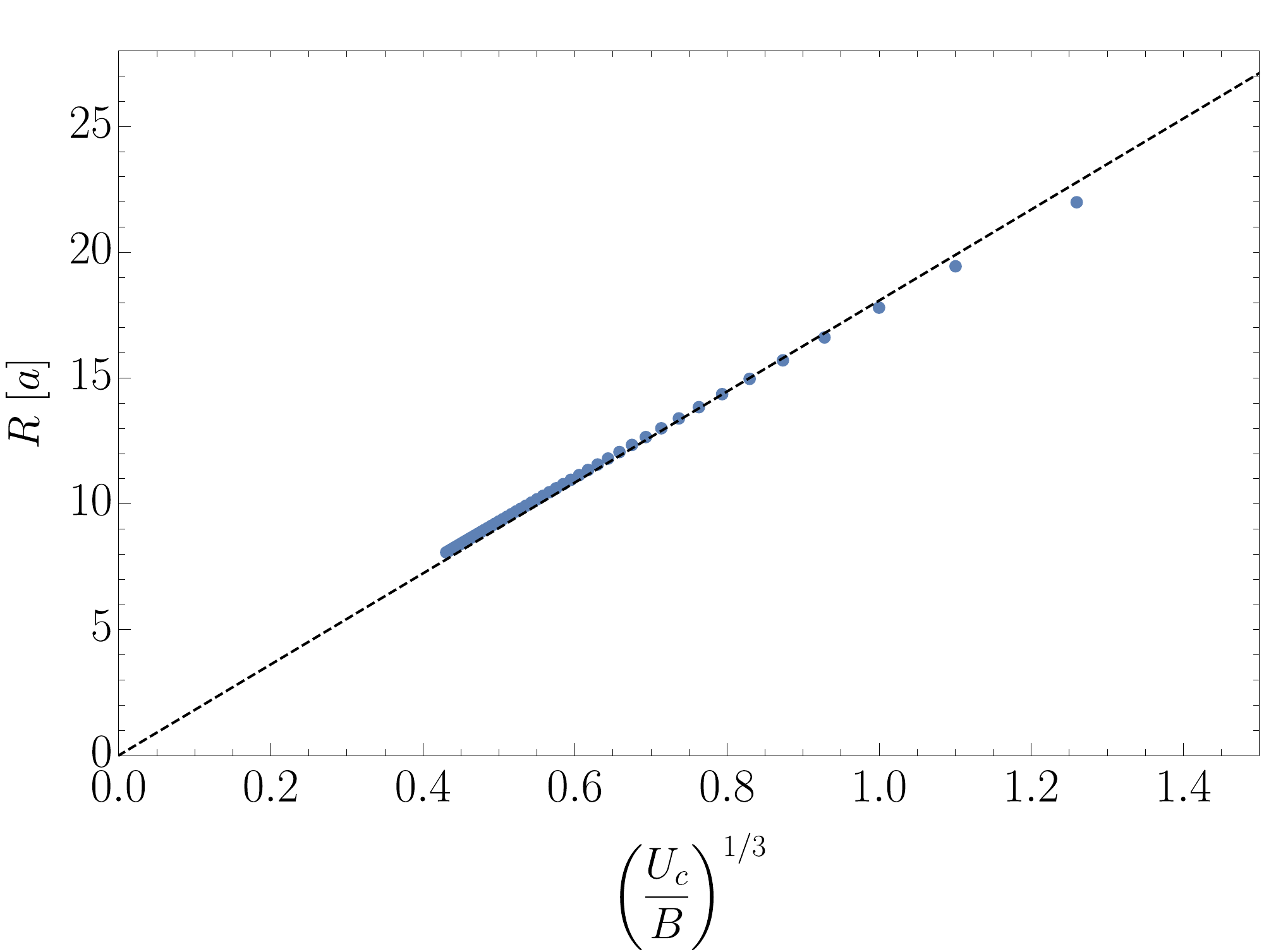}
	\caption{Skyrmion radius $R$ as a function of $\left(U_c/B\right)^{1/3}$ for $J=1$. The dashed line shows the fit in Eq.~\eqref{eq:RadiusFit}. Deviations for small radii happen due to discretization effects and at large $U_c/B$ because of the interactions between skyrmions.}
	\label{fig:radiusFitPowerLaw}
\end{figure}

To verify the formula for the skyrmion radius, Eq.~\eqref{eq:Radius}, we consider a low skyrmion density in our system and large enough magnetic fields, such that the skyrmions can be viewed as isolated and small. The skyrmion radius is defined by the distance from the skyrmion center to the point where the z-component of the magnetization is zero, meaning $m_z(R)=0$. For improved statistics we calculate the radius for four directions, namely the positive and negative x- and y-direction, and average the results. In Fig.~\ref{fig:radiusFitPowerLaw} the skyrmion radius as a function of $\left(U_c/B\right)^{1/3}$ is shown. Our numerics yields:
\begin{equation}
R = c_2 \left( \dfrac{U_c}{B}\right)^{1/3} \ \text{with} \ c_2 \approx 8.86 \ a \label{eq:RadiusFit}
\end{equation}
where $a$ is the discretization length of our numerics and $J=1$.
As the skyrmion size decreases, discretization effects become more important. This explains deviations for small $U_c/B$ from the expected behaviour in Fig.~\ref{fig:radiusFitPowerLaw}. Interactions between different skyrmions become more important when increasing  $U_c/B$ at finite winding number density. These interactions lead to a smaller value for $R$ than what is expected from Eq.~\eqref{eq:RadiusFit}.

\bibliography{bib_Coulomb}

\end{document}